\documentclass[twocolumn,5p,times]{article}
\usepackage{graphicx}
\usepackage{lscape,graphicx}
\usepackage{hhline}
\usepackage{multicol}
\usepackage{colordvi}
\usepackage{dcolumn}
\usepackage[fleqn]{amsmath}
\usepackage[mathscr]{euscript}
\usepackage[usenames, dvipsnames]{color}
\usepackage[usenames,dvipsnames,svgnames,table]{xcolor}
\usepackage[linktocpage]{hyperref}
\usepackage[a4paper,bindingoffset=0.2in,left=0.5in,right=0.5in,top=1.0in,bottom=1.0in,footskip=0.25in]{geometry}
\newcommand*{\doi}[1]{\href{http://dx.doi.org/#1}{doi: #1}}
\usepackage[numbers]{natbib}
\makeatletter
\renewcommand{\section}{\@startsection{section}{1}{\z@}{-3.5ex \@plus -1ex \@minus -.2ex}{1.3ex \@plus.2ex}{\normalfont\small\bfseries\boldmath}}
\makeatother

\makeatletter
\renewcommand{\subsection}{\@startsection{subsection}{2}{\z@}{-3.5ex \@plus -1ex \@minus -.2ex}{1.3ex \@plus.2ex}{\normalfont\small\bfseries\boldmath}}
\makeatother

\makeatletter
\renewcommand{\subsubsection}{\@startsection{subsubsection}{3}{\z@}{-3.5ex \@plus -1ex \@minus -.2ex}{1.3ex \@plus.2ex}{\normalfont\small\bfseries\boldmath}}
\makeatother

\makeatletter
\everymath{\if@display\else\thickmuskip=0mu plus 0mu\fi}
\makeatother

\title{\large \bf  {\tt ee$\in$MC}: Simulation of  $\bf e^{+}e^{-} \to Hadrons (n\gamma)$ Events }
\date{}

\author{\normalsize Ian M. Nugent$^{*}$ \\ \normalsize Victoria, B.C., Canada}

\begin{document}
\twocolumn[
  \begin{@twocolumnfalse}
    \maketitle
\begin{abstract}
The {\tt ee$\in$MC} generator package is extended to include $e^{+}e^{-} \to Hadrons (n\gamma)$ events where the photons can be generated in either the initial-state or final-state production.  
In particular, the $e^{+}e^{-} \to \pi^{+}\pi^{-} (n\gamma)$, $e^{+}e^{-} \to \Upsilon(4s) (m\gamma) \to B^{+}B^{-} (n\gamma)$ and $e^{+}e^{-} \to \Upsilon(4s) (m\gamma) \to B^{0}\bar{B}^{-} (n\gamma)$ 
decay channels have been implemented in the generator. The B-mesons are not decayed in  {\tt ee$\in$MC}, but are left for other generators that specialized in B-mesons. 
Both the initial-state-radiation and final-state-radiation are simulated using the Yennie-Frautschi-Suura Exponentiation procedure \cite{Yennie:1961}, where several effective models are implemented to investigate the
assumption of the radiative structure of the final-state vertex. The impact of these radiative models on the $e^{+}e^{-} \to \Upsilon(4s) (m\gamma) \to B^{+}B^{-} (n\gamma)$ and 
$e^{+}e^{-} \to \Upsilon(4s) (m\gamma) \to B^{0}\bar{B}^{-} (n\gamma)$  processes, excludes some models, particularly, in regards to the Coulomb potential. This has an important impact 
for meson and baryon systems near threshold.
\\ \\
Keywords: Electron-Positron Collider, Monte-Carlo Simulation \\ \\
\end{abstract}
\end{@twocolumnfalse}
]
%\begin{keyword}
%Tau Lepton \sep Electron-positron Collider \sep Monte-Carlo Simulation
%\end{keyword}

%\end{frontmatter}
%\linenumbers
\renewcommand{\thefootnote}{\fnsymbol{footnote}}
\footnotetext[1]{Corresponding Author \\ \indent   \ \ {\it Email:} inugent.physics@outlook.com}
\renewcommand{\thefootnote}{\arabic{footnote}}

\section{Introduction}

The $e^{+}e^{-} \to Hadrons (n\gamma)$ processes are central components of the physics programs at the B-Factories \cite{BaBar:1995bns,BaBar:1998yfb,abe2010belle,Kou_2019,BaBar:2014omp}. The quantum entangled states of 
B-mesons from which CP is 
studied are produced through the $\Upsilon(4s)$ resonance by means of the $e^{+}e^{-} \to \Upsilon(4s) (m\gamma) \to B^{+}B^{-} (n\gamma)$  and 
$e^{+}e^{-} \to \Upsilon(4s) (m\gamma) \to B^{0}\bar{B}^{0} (n\gamma)$ interactions \cite{BaBar:1995bns,BaBar:1998yfb,abe2010belle,Kou_2019,BaBar:2014omp}.  
The low energy $e^{+}e^{-} \to Hadrons (n\gamma)$ processes are  an essential
element in the determination of the hadronic vacuum polarization contribution to the $g-2$ anomaly \cite{Gourdin:1969dm} and are
related to the hadron production in weak interactions in $\tau$ decays through the Conserved-Vector-Current hypothesis 
\cite{Halzen:1984mc} for the vector current. 
The addition of the $e^{+}e^{-} \to Hadrons (n\gamma)$ into {\tt ee$\in$MC} allows for a consistent frame-work for investigating the low energy hadronic systems in  
the $e^{+}e^{-} \to Hadrons (n\gamma)$ and hadronic  $\tau$ decay processes. 
The $e^{+}e^{-} \to Hadrons (n\gamma)$ processes are implemented within the Yennie-Frautschi-Suura (YFS) Exponentiation Formalism \cite{Yennie:1961} for the subtraction of the infrared QED divergence 
\cite{Bloch:1937},
following the convention applied to $e^{+}e^{-}\to \mu^{+}\mu^{-}(n\gamma)$ and  $e^{+}e^{-}\to \tau^{+}\tau^{-}(n\gamma)$ \cite{Nugent:2022ayu}.
In contrast to the $e^{+}e^{-}\to \mu^{+}\mu^{-}(n\gamma)$ and  $e^{+}e^{-}\to \tau^{+}\tau^{-}(n\gamma)$ the final-state particles are ``composite particles'' and therefore, the description of the 
radiative emissions depends on the sub-structure. The formalism for implementing the  $e^{+}e^{-} \to Hadrons (n\gamma)$ processes are described in Section \ref{sec:Form}, while
in Section \ref{sec:YFS}, the effective models for the radiative emission within the YFS Formalism are described. 
The hadronic models for  $e^{+}e^{-} \to \pi^{+}\pi^{-} (n\gamma)$ and
 $e^{+}e^{-} \to \Upsilon(4s) (m\gamma) \to B^{+}B^{-} (n\gamma)$  and
$e^{+}e^{-} \to \Upsilon(4s) (m\gamma) \to B^{0}\bar{B}^{0} (n\gamma)$  are described in Sections \ref{sec:pipi} and \ref{sec:Y4s}.
The impact of
the phase-space constraints of the  $e^{+}e^{-} \to \Upsilon(4s) (m\gamma) \to B^{0}\bar{B}^{-} (n\gamma)$  and $e^{+}e^{-} \to \Upsilon(4s) (m\gamma) \to B^{0}\bar{B}^{-} (n\gamma)$ processes are exploited
to discriminate between the effective radiative emission models in Section \ref{sec:Y4s}.

\section{$e^{+}e^{-}\to Hadrons$ Formalism \label{sec:Form}}

The cross-section for $e^{+}e^{-}\to Hadrons$ can be described within the YFS Exponentiation procedure \cite{Yennie:1961} as:

\begin{equation}
\resizebox{0.375\textwidth}{!}{$
d\sigma= \frac{\sum_{a=0}^{\infty} Y_{i}(Q_{i}^{2})Y_{f}(Q_{f}^{2})|\sum_{b=1}^{\infty}\bar{{\mathcal M_{a}^{b}}}|^{2} dPS_{a}^{\delta M}}{4(|\vec{P}_{e^{-}}|E_{e^{+}}+E_{e^{-}}|\vec{P}_{e^{+}}|)}
$}
\label{eq:YFS}
\end{equation}

\noindent where ${\mathcal M_{a}^{b}}$ is the sum of diagrams that contribute to the matrix element for $a$ hard photons and $b$ internal photon lines between initial and final-state, $Y_{i}(Q_{i}^{2})$ 
is the initial-state YFS Exponential Form Factors described in \cite{Nugent:2022ayu} and $Y_{f}(Q_{f}^{2})$ is the corresponding final-state radiative model. 
The final-state radiative models will be discussed in Section \ref{sec:YFS}.% and their implications will be discussed in Section \ref{sec:Y4s}. 
The hard final-state radiative photon in the matrix element ${\mathcal M_{a}^{b}}$ 
is implemented using the Feynman rules for the spin-0 photon vertices and is embedded in the hadronic current.  
The matrix elements included in {\it ee$\in$MC} for $e^{+}e^{-}\to Hadrons$  are Born level ($\bar{{\mathcal M}}_{a=0}^{b=1}$),  LO ($\bar{{\mathcal M}}_{a=1}^{b=1}$)
NLO ($\bar{{\mathcal M}}_{a=2}^{b=1}$), NNLO ($\bar{{\mathcal M}}_{a=3}^{b=1}$) 
in conjunction with the initial-state NNNLO ($\bar{{\mathcal M}}_{a=4}^{b=1}$).
The Born level matrix element for a  $e^{+}e^{-}\to V\to PP$ hadronic interaction 
may be written as:

\begin{equation}
\resizebox{0.400\textwidth}{!}{$
\begin{array}{lll}
{\mathcal M_{a=0}^{b=1} }&=&\left(\bar{v}_{2}\imath e \gamma^{\mu}u_{1}\right)\frac{g_{\mu \nu}}{q^{2}} \imath e Q_{f} N_{c} \frac{\left(g^{\nu\alpha}-q^{\nu}q^{\alpha}\right)}{P(s)} A(s) (p_{a}-p_{b})_{\alpha}  \\
&=&\left(\bar{v}_{2}\imath e \gamma^{\mu}u_{1}\right)\frac{g_{\mu \nu}}{q^{2}}\imath e Q_{f} J^{\nu}
\end{array}
$}
\label{eq:ME}
\end{equation}

\noindent
where $Q_{f}$ is the quark charge and $J^{\nu}$ is the hadronic current implemented within the framework of a given model. The propagator $P(s)$ and decay amplitude $A(s)$ are some functions contained in the
hadronic current, $J^{\nu}$.   This generic formalism with the current is employed to allow for  
interchanging the hadronic currents between the $\tau$ decays and production in $e^{+}e^{-}$ interactions. Within the Form-Factor approach for the Formal-Vertex-Structure \cite{Peskin:1995ev}, the hadronic
current for the two body $V\to PP$ decay may be written as $J^{\nu}=N F(s)\left(g^{\nu\alpha}-q^{\nu}q^{\alpha}\right)(p_{a}-p_{b})_{\alpha}$, where $F(s)$ is the Form-Factor, 
$\left(g^{\nu\alpha}-q^{\nu}q^{\alpha}\right)$ is the tensor component of the massive spin 1 propagator,  $(p_{a}-p_{b})^{\alpha}$ is the 
spin-1 spin-0 Feynman vertex coupling, $p_{a}$ and $p_{b}$ are the outgoing mesons and $N$ is the current amplitude for the hadronic model.
For models where the amplitude of the hadronic current, $N$, is constructed from the $SU(3)$ group generators through an effective 
Lagrangian \cite{Georgi:2009} the colour factor $N_{c}$ is implicitly included \cite{Peskin:1995ev}. Therefore, we use the convention that the colour factor $N_{c}$ is omitted in the second part of Eq. 
\ref{eq:ME}, but is instead included in the hadronic current amplitude. The spin formalism is implemented using the algorithm in \cite{Nugent:2022dni}, and therefore allows for an arbitrary initial spin state
for the incoming $e^{+}e^{-}$ pair.

\section{Exponentiation and Radiative Models \label{sec:YFS}}

In contrast to the $e^{+}e^{-}\to \mu^{+}\mu^{-}(n\gamma)$ and  $e^{+}e^{-}\to \tau^{+}\tau^{-}(n\gamma)$ interactions the final-state particles are ``composite particles'' and therefore, 
the description of the radiation must be parameterized within an effective model for the sub-structure. In \cite{BaBar:2015onb}, two fundamental descriptions of the possible effective models for
describing the final-state radiation are defined in terms of Klein-Gordon and Dirac spinor states.  However, the description of the QED interaction at the final-state production vertex must include both the 
exponentiation with the radiative emission model as well as the Coulomb potential. Therefore, the possible effective models for the radiative emissions are:

\begin{itemize}
\item Model 1: The QED time scale is assumed to be much greater than the QCD scale for forming the meson and therefore the outgoing particles at the vertex are assumed to be radiating at the meson level and are
treated as spin-0 particles in the exponentiation and radiative emission \cite{BaBar:2015onb} and as the fully formed mesons for the Coulomb potential. The YSF Exponential Form-Factor is determined from the 
soft and 
virtual contributions for the spin-0 particles \cite{Schwinger:1998}  by relating these terms to the  ${\mathcal O}(\alpha)$ term in the exponential series \cite{Peskin:1995ev}
in an analogous procedure to Type-IV Exponentiation in \cite{Nugent:2022ayu}. The Coulomb potential is factorized out into a separate series  \cite{Nugent:2022ayu}, the Sommerfeld-Sakharov factor 
\cite{Smith1994117}, and is formulated using kinematics of the fully formed meson. The exponential factor can then be expressed as:

\begin{equation}
\resizebox{0.4\textwidth}{!}{$
\begin{array}{ll}
Y_{spin-0}=&e^{\frac{2\alpha Q_{x}^{2}}{\pi}\left(\left[\left(1+v^{2}\right)\chi(v)-1\right]\left[\ln\left(\frac{\delta M}{m_{l}}\right)+1\right]+\chi(v)-\frac{(1+v)}{v}\int_{0}^{v}\frac{dv'\chi(v')}{1-v'^{2}} \right)} \times \\
&e^{-\frac{\alpha Q_{x}^{2}}{\pi}\left(P\int_{0}^{1}\frac{dv'\left(1+v'^{2}\right)\ln\frac{v'^{2}}{1-v'^{2}}}{v^{2}-v'^{2}}\right)-F_{c}|_{{\mathcal O}(\alpha)}}
\end{array}
$}
\end{equation}
\noindent where $\chi(v)=\frac{1}{2v}\ln\left(\frac{1+v}{1-v}\right)$, $\delta M=\sqrt{s+E_{soft}^{2}}-\sqrt{s}+E_{soft}$ and $Q_{x}$ is the charge of the outgoing particle $x$.

\item Model 2: The hadron formation is based on the assumption that colour singlets must be formed between each of the valence quarks and one of the newly created quarks before the 
meson or baryon can disintegrate\footnote{The picture given in this model is inspired by the Quark Model \cite{Halzen:1984mc}, Flux-Tube-Breaking Model 
\cite{Isgur:1988vm,Godfrey:1985xj,Kokoski:1985is,Isgur:1983wj,Isgur:1984bm} and the Quark-Pair-Creation Model \cite{LeYaouanc:1972vsx,LeYaouanc:1977fsz,Ono:1980js}.}. Therefore, the 
decay process can be 
described by a two-step process. Firstly, the colour singlets are formed through the ``flux-tube'' between 
the valence quarks ``breaking'' and generating a quark-anti-quark pair which firmly attaches each valence quark to each one of the newly created quarks. The quark pairs are tightly bound 
through a new pair of ``flux-tubes'' creating two colour singlets.  
%The ``flux-tube'' breaking and formation of the new colour singlet pair is assumed to occur before the on a time scale substantially less that the QED time scale. 
%This step begins the disintegration of the meson or baryon. 
At this stage, the colour singlets are not stable particles nor independent of each other,
but bound through a residual QCD potential.
In the second step, the colour singlets develop into the final-state mesons
with the given angular configuration through the residual QCD force interacting with the tightly bound colour singlets. It is assumed that this interaction proceeds at a time-scale long enough for the QED
force to interact with the colour singlets through the Coulomb potential, a consequence of the Coulomb potential being dominated by smaller time-scales.
At this time-scale in the decay process, the charge of the colour singlet corresponds to that of the final-state meson, however, the mass and kinematics correspond to that of the quark level colour singlet, 
where we approximate the effective mass of the colour singlet as $m_{b}+m_{q-light}$.
This picture only presumes that the total angular momentum, of the colour singlets and gluon field is constrained 
by the conservation of momentum for the total system.
For simplicity in the two-body $V\to PP$ interaction, we assume the gluon field carries the spin 1 angular momentum for most of this time-scale. This assumption is based on a rapid convergence of
the spin configuration to the final-state spin configuration through the strong interaction. However, at the initial formation of the colour singlets, the quark pairs could be in an integer spin configuration 
that then radiates to a spin-0 state. The latter assumption assumes that the radiative emission through QED is supressed relative to the QCD interaction due to the relative strengths of the coupling 
constants\footnote{This assumption could be tested experimentally.}, $\alpha$ and $\alpha_{s}$. 
Under this assumption the intermediate state colour singlets and final-state mesons can be treated as spin 0 particles in this 
model\footnote{The wave-length of the radiative photons 
is $\gg$ than the size of the colour singlets, and therefore one can approximate the colour singlet as a point particle.}. Therefore, YSF Exponential Form-Factor is 
determined from the soft-photon and virtual contributions for the spin-0
particles \cite{Schwinger:1998} through the same procedure as Model 1 using the final-state mesons for the effective average mass over the time-scale of the QED radiative 
process.
This picture of colour singlets being $\ll$ than the radiative wave-length of the QED emissions also justifies the negligible radiative emission directly from the $\Upsilon(4s)$.
%\footnote{This assumption assumes $\tau_{QED}>>\tau_{QCD}$ for the radiative process. If this assumption is not made a lower effective mass somewhere between the meson mass and 
%singlet $q\bar{q}$ mass would be used.}, 
%under the assumption asymptotic formation of the mesons. 
 
\item Model 3: The hadron formation is assumed to occur rapidly $\tau_{QED}\gg\tau_{QCD}$ that the radiative QED processes occur at meson level, however, it is sufficiently slow that the 
Coulomb potential is dominated by the quark level structure. Again, this implies that the YSF Exponential Form-Factor corresponds to the spin-0 Form-Factor from Model 2. However, the Coulomb potential
is determined through the valence quarks charge and kinematics \footnote{In B-mesons produced from the $\Upsilon(4s)$ resonance, the contribution of the light quarks are negligible do to the high relative 
boost.}.           

\item Model 4: The outgoing particles at the vertex are assumed to radiate at quark level and are treated as Dirac Spinors in the exponentiation along with the Coulomb potential \cite{BaBar:2015onb}.
 The YSF Exponential Form-Factors are constructed using Exponentiation Type-IV from \cite{Nugent:2022ayu}.
\end{itemize}

\noindent Models 3 and 4 are logical extensions of the quark level radiative Model in \cite{BaBar:2015onb}. Based on confinement, we argue Models 1 and 2 are more likely\footnote{Model 4, 
is not explicitly implemented in the generator, however it has been included for consistency with \cite{BaBar:2015onb}.}\footnote{This is consistent with the observations in \cite{BaBar:2015onb}.}.

\section{$e^{+}e^{-}\to \pi^{-}\pi^{+}(n\gamma)$ Model \label{sec:pipi}}

\begin{figure*}[tbp]
  \begin{center}
    \resizebox{260pt}{185pt}{
      \includegraphics{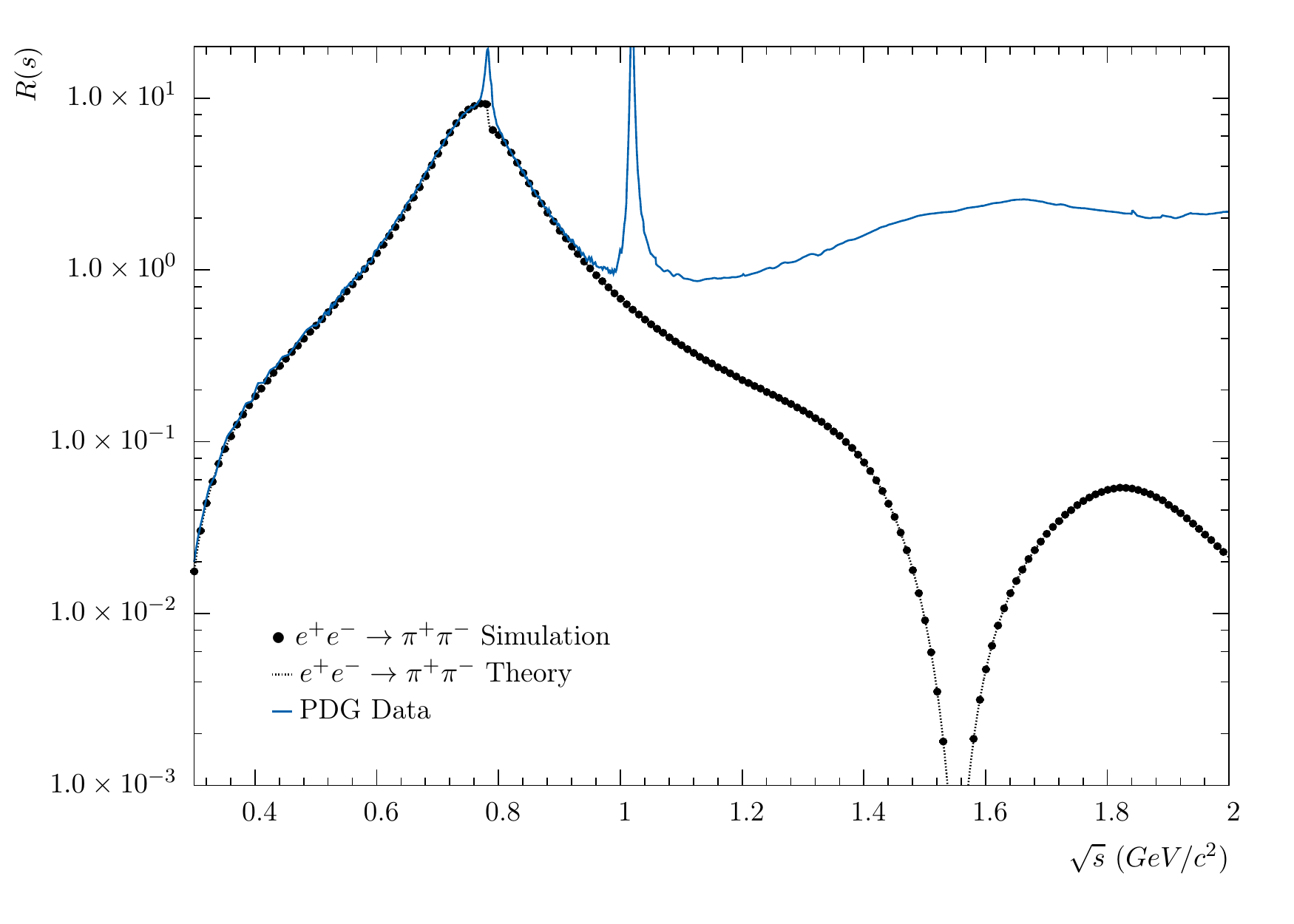}
    }
    \caption{ $R(s)$ as a function of the collisions energy ($\sqrt{s}$) for the total $R(s)$ from world average \cite{PDG2020} (blue-line) overlayed with the $\pi^{+}\pi^{-}$ contributions
determine using a scan of the Born level simulation of $e^{+}e^{-}\to\pi^{+}\pi^{-}$ events (black-points) and theory (black-line). 
 \label{Rhad_pipi}}
  \end{center}
\end{figure*}

The hadronic current in the $e^{+}e^{-}\to \pi^{-}\pi^{+}(n\gamma)$ is described with the vector dominance based Gounaris-Sakurai \cite{Gounaris:1968mw} Model using the formalism 
from \cite{PhysRevD.86.032013}. The normalization of the current is obtained from the Chiral-Lagrangian \cite{Georgi:2009}. An overlay of the $R(s)$ simulated for the  $e^{+}e^{-}\to \pi^{-}\pi^{+}$
interaction on the world average $R(s)$ data \cite{PDG2020}  can be seen in Figure \ref{Rhad_pipi}.
The model parameters have been tuned to improve agreement with world average $R(s)$ data \cite{PDG2020}. 
Of particular interest is the theoretical uncertainty on the $e^{+}e^{-}\to \pi^{-}\pi^{+}(n\gamma)$ prediction, more specifically the uncertainty due to the truncated terms in the perturbative Feynman series. 
The convergence of the cross-section is particularly sensitive to the initial-state owing to the small mass of the incoming electron and positron. Thus, for the soft-photon cut-off range recommended  
for the B-Factory energies, the associated truncation uncertainty on the perturbative Feynman series due to the soft-photon cut-off is similar to 
$e^{+}e^{-}\to \mu^{-}\mu^{+}(n\gamma)$ in \cite{Nugent:2022hse}, at 
the ${\mathcal O}(2-3\%)$ level.
However, due to the identical treatment of the initial-state
radiation between the $e^{+}e^{-}\to \pi^{-}\pi^{+}(n\gamma)$ and  $e^{+}e^{-}\to \mu^{-}\mu^{+}(n\gamma)$ interactions, the theoretical uncertainty from the initial-state radiation will mostly cancel
in a ratio except the small differences from phase-space. The truncation uncertainty in the   $\sigma(e^{+}e^{-}\to \pi^{-}\pi^{+}(n\gamma))/\sigma(e^{+}e^{-}\to \mu^{-}\mu^{+}(n\gamma))$ ratio
would therefore primarily come from the final-state radiation and could be estimated with the procedure in \cite{Nugent:2022hse}. Within the theoretical truncation uncertainty, such an observable can be treated as
infrared safe, including if there are selection criteria applied to the number of photons \cite{Nugent:2022hse}.  Naively, this would be $<1\%$.
For the low energy $\pi^{+}\pi^{-}$ mass range produced through a hard initial-state-radiative emission in the $e^{+}e^{-}\to \pi^{-}\pi^{+}(n\gamma)$ process, 
the uncertainty will be greater since the radiative tail from initial-state-radiation converges more slowly than the total cross-section which is discussed in \cite{Nugent:2022hse}.
  The addition of higher order terms in the perturbative Feynman series are expected to be required in the $e^{+}e^{-}\to \pi^{-}\pi^{+}(n\gamma)$ interaction
before the theoretical precision is sufficient for an analysis using the Initial-State Radiation Method \cite{Binner:1999bt} at the B-Factories.

\section{$e^{+}e^{-}\to \Upsilon(4s) (m\gamma) \to  BB(n\gamma)$ Model \label{sec:Y4s}}

\begin{figure*}[tbp]
  \begin{center}
    \resizebox{260pt}{185pt}{
      \includegraphics{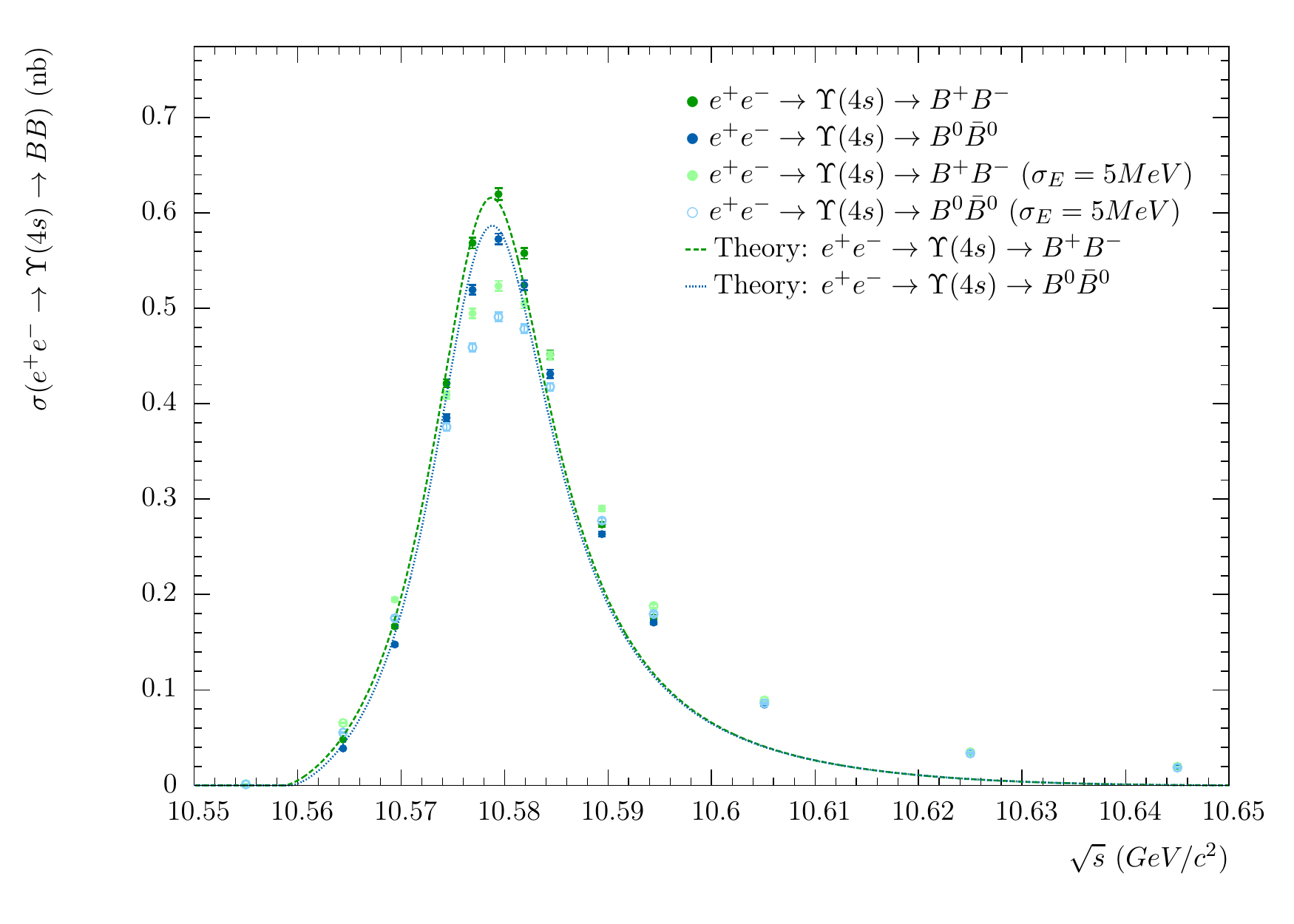}
    }
    \caption{ The line-shape of the $\Upsilon(4s)$ resonance produced through the $e^{+}e^{-}\to\Upsilon(4s)\to B^{+}B^{-}$
and $e^{+}e^{-}\to\Upsilon(4s)\to B^{0}\bar{B}^{0}$  interactions. The initial-state radiation is based on Type-IV Exponentiation while the final-state-radiation is based on Model 2 with a
soft-photon cut-off of $1MeV$. The solid points
correspond to the simulated production cross-section at the given $\sqrt{s}$ where the open points correspond to the simulated cross-section taking into account the beam-spread, $\sigma=5MeV$, through numerical
 integration for  $e^{+}e^{-}\to\Upsilon(4s)\to B^{+}B^{-}$ (green)
and $e^{+}e^{-}\to\Upsilon(4s)\to B^{0}\bar{B}^{0}$ (blue) interactions. The curves are the analytical Born level theoretical computation
 taking into account the phase-space,
$\sigma=\frac{\pi\alpha^{2}(s)\beta^{3}Q_{f}^{2}N^{2}|F(s)|^{2}}{3s}$ where $|F(s)|^{2}  \propto \frac{\Gamma_{B_{x}\bar{B}_{y}}(s)\times (s/\beta^{3}) }{(s-\bar{m}^{2}(s))^{2}+M^{2}\Gamma_{total}^{2}(s)}$ \cite{PhysRevD72.032005,ARGUS:1994lbi}.
To account for the radiative corrections from the YFS Exponentiation procedure the average cross-section of the theoretical curves are normalized to the
average simulated $e^{+}e^{-}\to\Upsilon(4s)\to B^{+}B^{-}$ and $e^{+}e^{-}\to\Upsilon(4s)\to B^{0}\bar{B}^{0}$  cross-sections at the pole-mass. This allows for the difference in the phase-space to be included. The results
are consistent with expectations \cite{BaBar:1995bns,BaBar:1998yfb}.
 \label{fig:Y4sscan}}
  \end{center}
\end{figure*}

\begin{figure*}[tbp]
\begin{center}
  \resizebox{520pt}{185pt}{
    \includegraphics{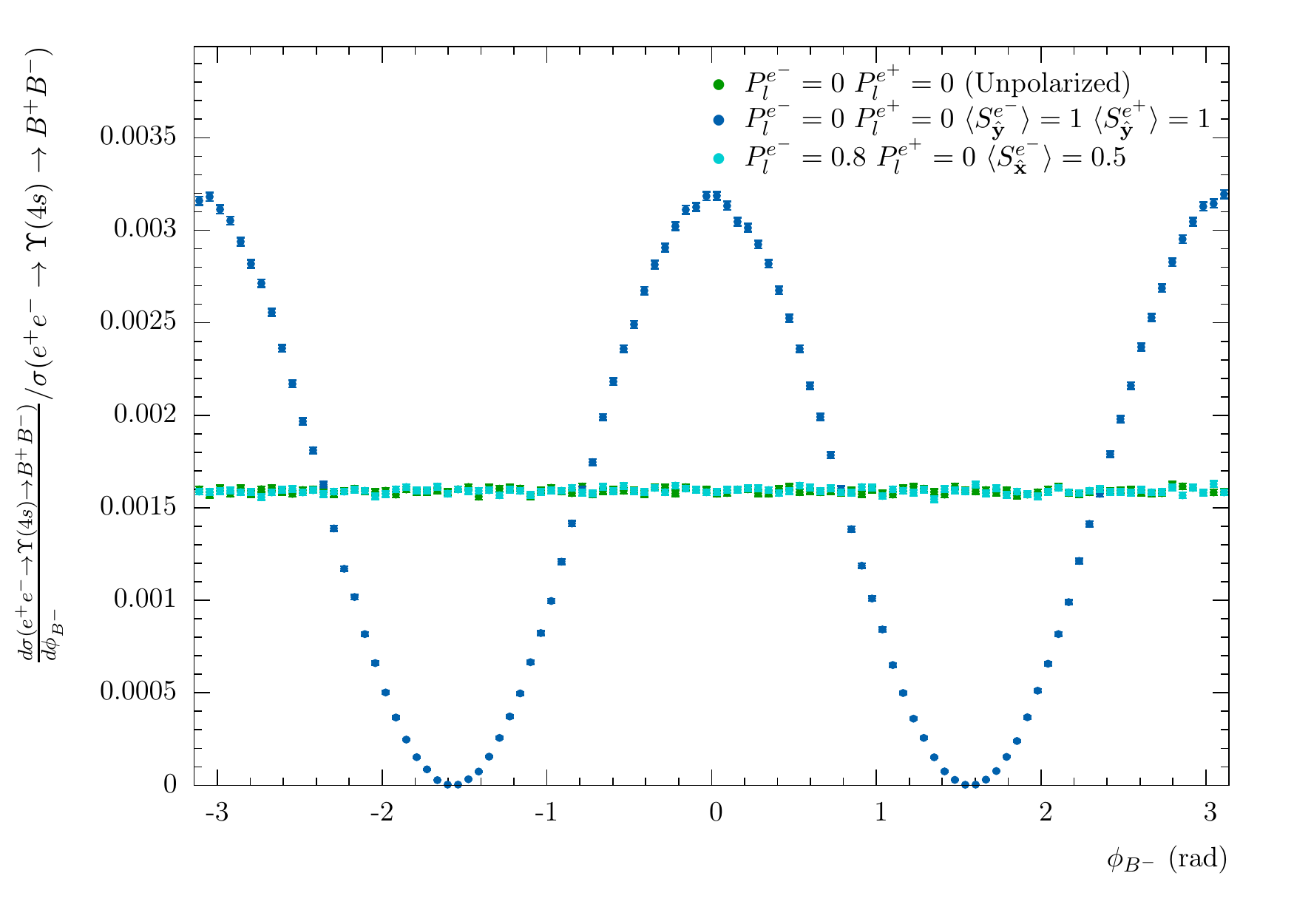}
    \includegraphics{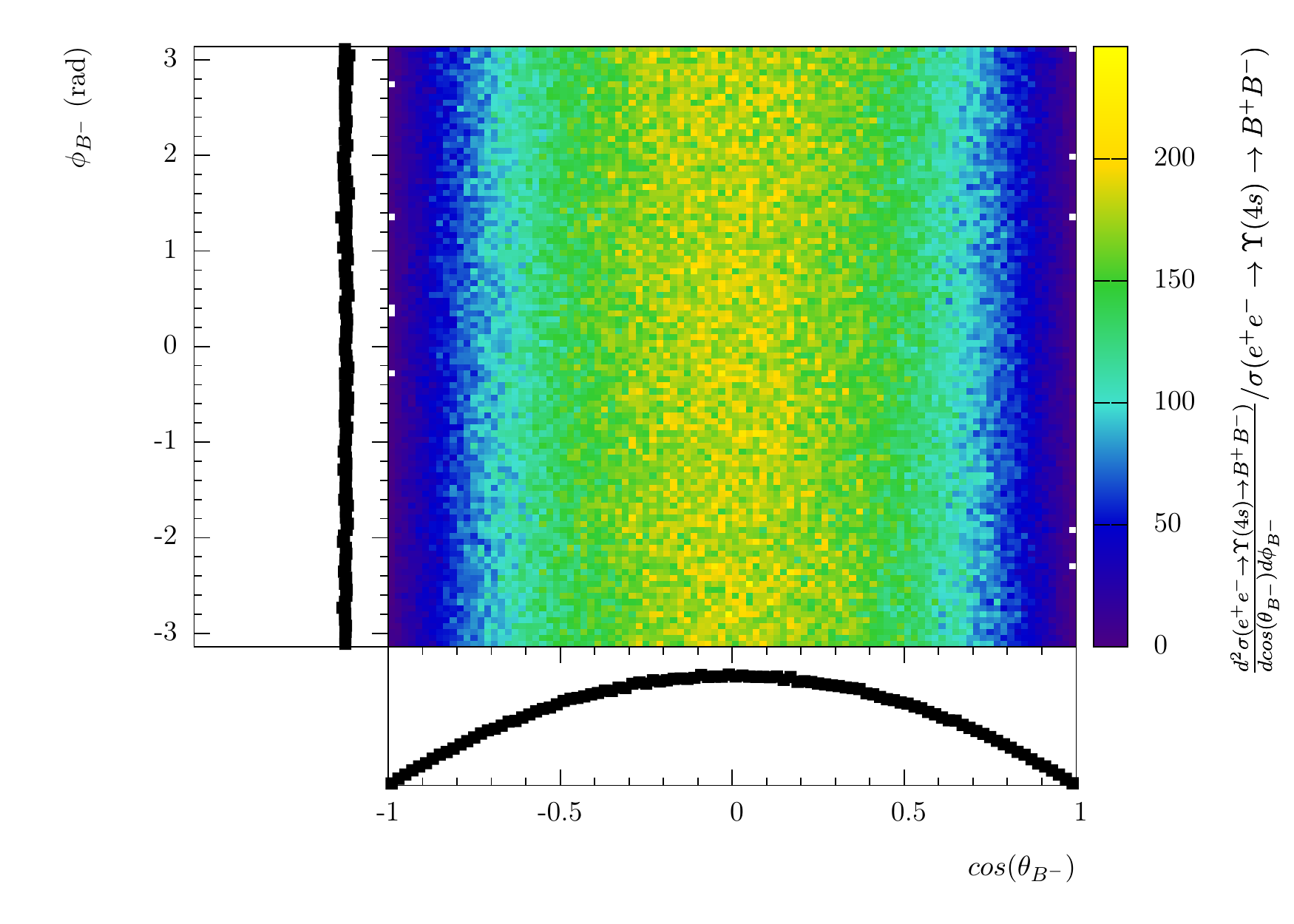}
  }
\resizebox{520pt}{185pt}{
    \includegraphics{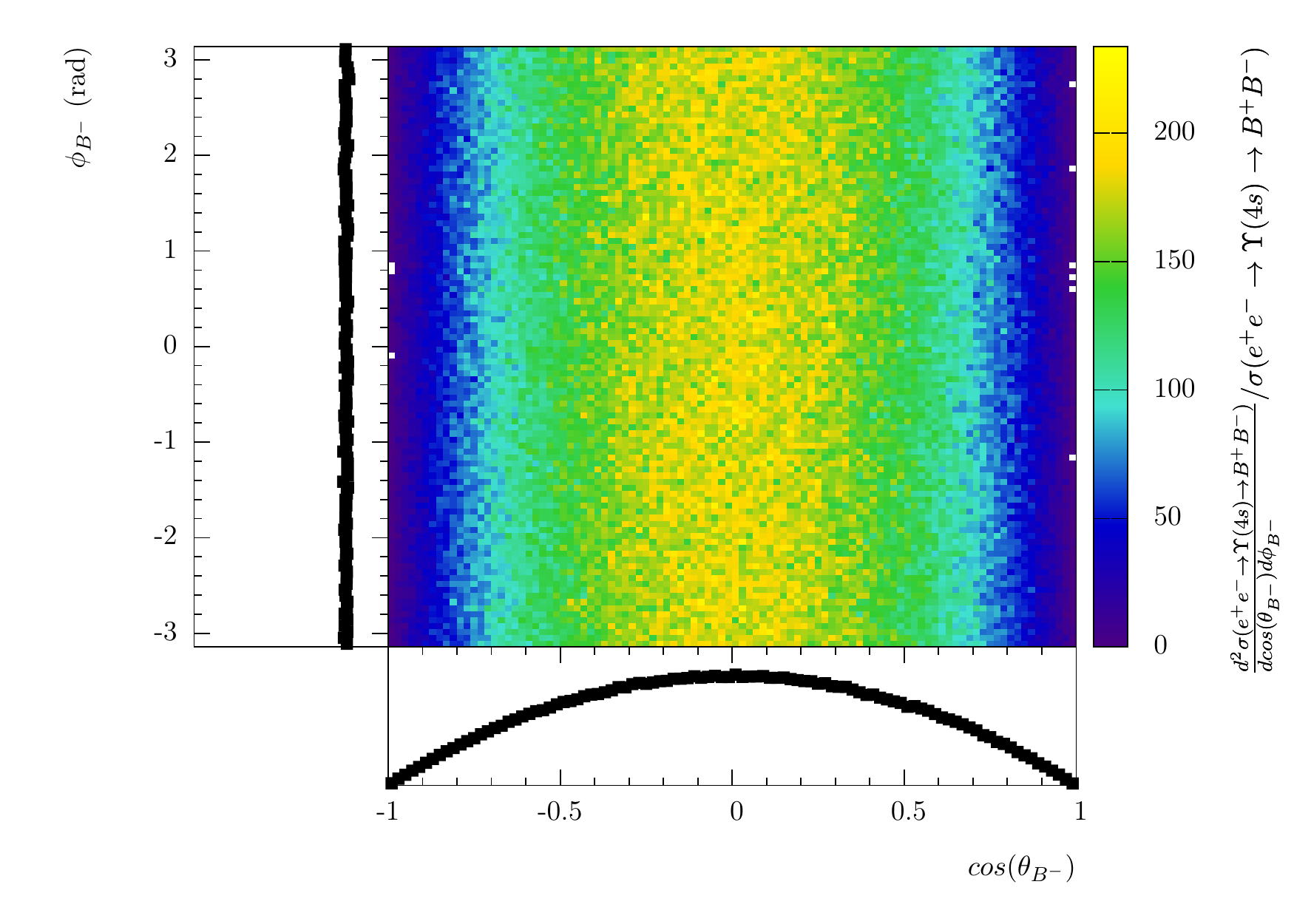}
    \includegraphics{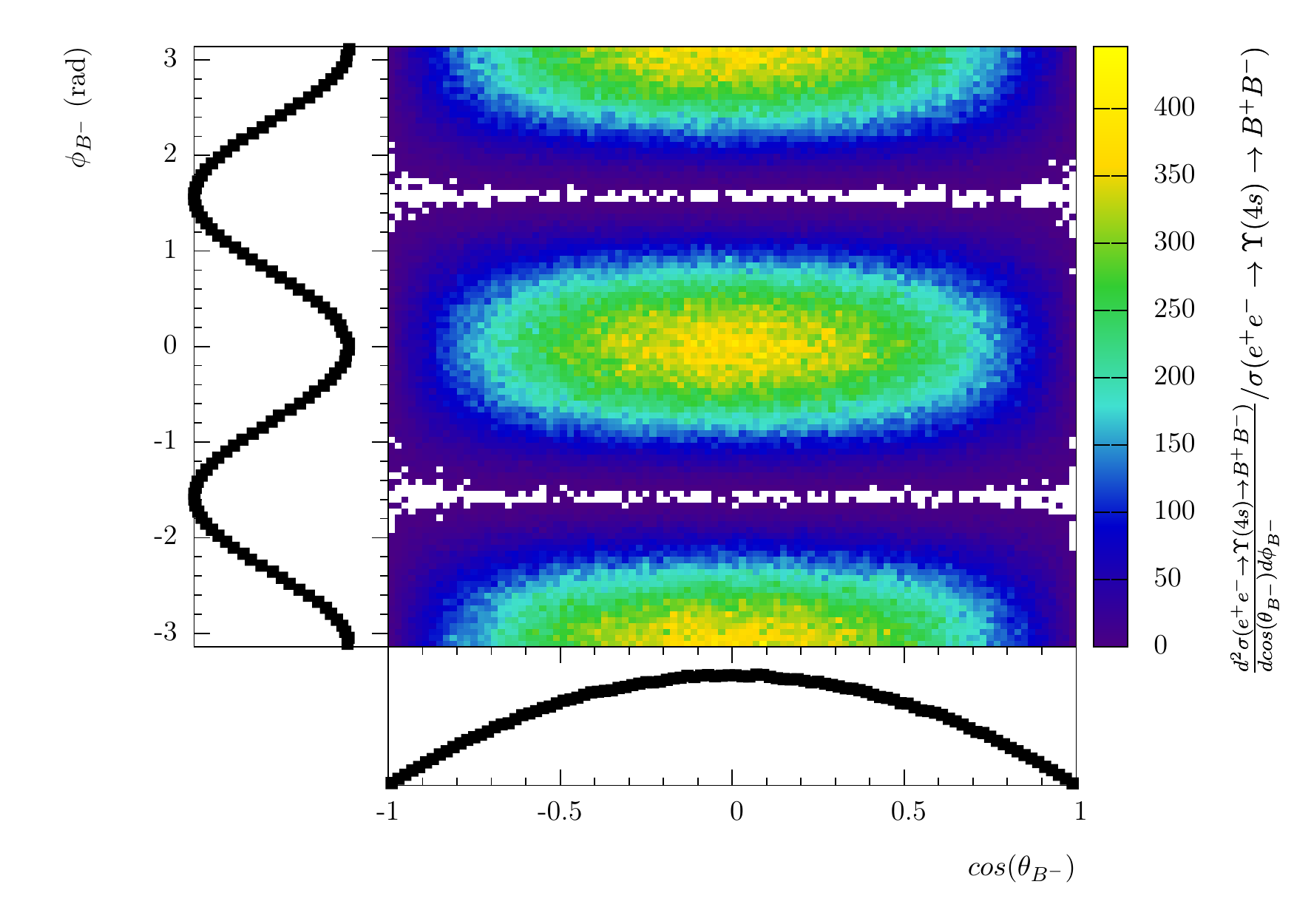}
  }
  \caption{The $\phi_{B^{-}}$ dependence of B-meson distribution for unpolarized, 100\% polarized along the y-axis and for a realistic polarization production scenario at the proposed polarization
upgrade at Belle-II \cite{USBELLEIIGROUP:2022qro}   (upper-left). The full $cos(\theta_{B^{-}})$ and $\phi_{B^{-}}$ dependence on the initial-state polarization for unpolarized initial-state (lower-left),
 100\% transversely polarized initial-state along the y-axis (lower-right) and the  realistic polarization scenario at the proposed polarization
upgrade at Belle-II \cite{USBELLEIIGROUP:2022qro} (upper-right) simulated at Born level.  \label{fig:spin}}
\end{center}
\end{figure*}

\begin{figure*}[tbp]
  \begin{center}
    \resizebox{260pt}{185pt}{
      \includegraphics{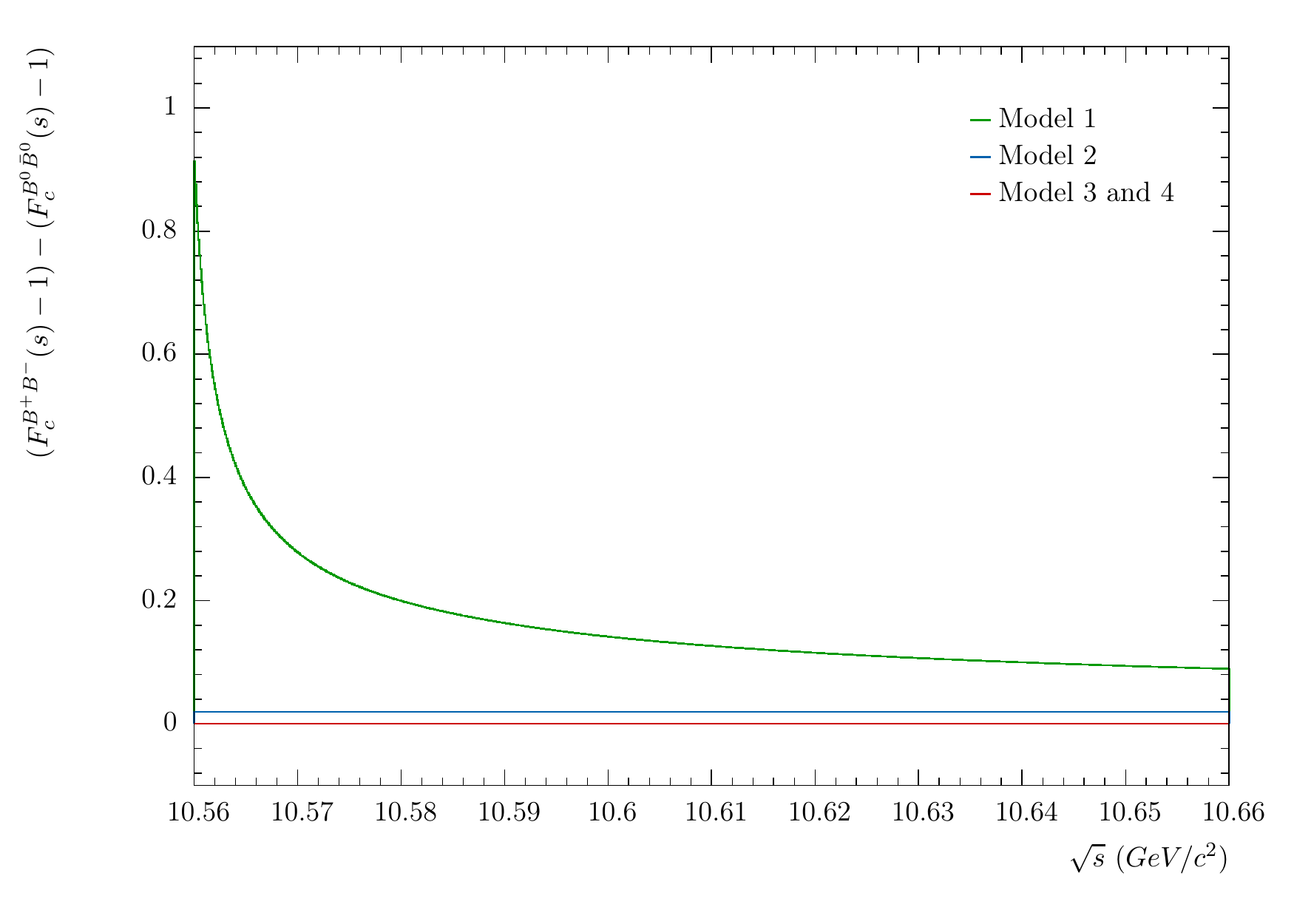}
    }
    \caption{ The difference in the relative amplification of the $e^{+}e^{-}\to\Upsilon(4s)\to BB$ cross-section from the Coulomb potential term $F_{c}$ between the $e^{+}e^{-}\to\Upsilon(4s)\to B^{+}B^{-}$ 
and $e^{+}e^{-}\to\Upsilon(4s)\to B^{0}\bar{B}^{0}$  decay modes for the exponential models presented in Section \ref{sec:YFS}.
 \label{fig:Fc}}
  \end{center}
\end{figure*}

The $e^{+}e^{-}\to \Upsilon(4s) (m\gamma) \to  B^{+}B^{-}(n\gamma)$ and  $e^{+}e^{-}\to \Upsilon(4s) (m\gamma) \to  B^{0}\bar{B}^{0}(n\gamma)$ interaction are implemented within the 
Quark-Pair-Creation (QPC) Model \cite{LeYaouanc:1972vsx,LeYaouanc:1977fsz,Ono:1980js}, where the Form-Factor is constructed from the QPC Model following the convention in 
\cite{PhysRevD72.032005,ARGUS:1994lbi} using dispersion relations for the mass shift function \cite{ARGUS:1994lbi}. The Form-Factor from the QPC Model is normalized within the Chiral-limit with
$\Gamma\ll M$, 
$F(s)\to\frac{M^{2}}{(s-\bar{m}^{2}(s))+\imath M\Gamma_{total}(s)}$, to enable the current amplitude normalization to be determined from the effective Chiral Lagrangian.
The current amplitude normalization can then be determined in the $SU(3)$ chiral limit for the light quarks 
where the heavy quarks transform as singlets 
\cite{Georgi:1999}. The effective Lagrangian is constructed from only the light quarks/anti-quarks in the heavy quark mesons which transform under $3$ and $\bar{3}$ \cite{Georgi:2009,Georgi:1999}.   
Within the QPC Model, the hadronic widths for B-meson pair production, $\Gamma_{B_{x}\bar{B}_{y}}(s)$, depends on the harmonic-oscillator wave-function and are not simply the phase-space and $s$ dependence 
factor\footnote{The it can be seen in Eq. 2-5 from \cite{PhysRevD72.032005} that the hadronic widths for B-meson pair productions are proportional to $\beta^{3}/s$.}, therefore the kinematic factors from
Eq. \ref{eq:ME} must be transformed $\beta^{3}/s\to \Gamma_{B_{x}\bar{B}_{Y}}(s)$ using the pole-mass for the normalization. The correction is included within the Form-Factor and results in a 
cross-section proportional to
$ \frac{\Gamma_{B_{x}\bar{B}_{y}}(s)\times (s/\beta^{3}) }{(s-\bar{m}^{2}(s))^{2}+M^{2}\Gamma_{total}^{2}(s)}$.
The relative amplitudes of the $ \Upsilon(4s) \to B^{+}B^{-}$ and  $\Upsilon(4s) \to B^{0}\bar{B}^{0}$ in the  $SU(3)$ limit are $1/\sqrt(2):1/\sqrt(2)$. As a consequence, at $M_{\Upsilon(4s)}$, the
relative fraction of $ \Upsilon(4s) \to B^{+}B^{-}$ to  $\Upsilon(4s) \to B^{0}\bar{B}^{0}$ is $(51.21)\%:(48.79)\%$ due to phase-space constraints ($\beta_{B^{+}B^{-}}^{3}/\beta_{B^{0}\bar{B}^{0}}^{3}$) and
 $(51.9\pm0.52)\%:(48.03\pm0.48)\%$ in the simulated prediction with Exponentiation Type-IV for the initial-state radiation and Model 2 for the final-state-radiation with a soft-photon cut-off of $1MeV$. 
These are both consistent with the average branching fractions $(51.4\pm0.6)\%:(48.6\pm0.6)\%$ \cite{PDG2020}.
The B-mesons are pseudo-scalar particles and therefore can be decayed without additional information by generators that specialize in $B$ decays. 
A line scan of $e^{+}e^{-} \to \Upsilon(4s) (m\gamma) \to BB(n\gamma)$ using the preferred final-state-radiation Model 2 is shown in Figure \ref{fig:Y4sscan} for the $ B^{+}B^{-}$ threshold up to the 
first node in the  $ \Upsilon(4s)$ width, with YFS Exponentiation and matrix elements $\bar{{\mathcal M}}_{a}^{b=1}$ where $a=0,1,2,3$ with and without a beam-spread of $5MeV$. 
Higher orders of radiation are not included, because the perturbative series at the pole-mass has converged in expansions of $\bar{{\mathcal M}}_{a}^{b=1}$ sufficiently due to the limited phase-space 
availability\footnote{Although a full check of the soft-photon cut-off
can not be applied as in \cite{Nugent:2022hse}, due to the narrow width of the resonance ($\delta M \ll \Gamma$), the cross-section at the  $\Upsilon(4s)$ pole mass is marginally consistent for 
soft-photon cut-offs between 1-5MeV in samples generated with a $0.3\%$ 
statistical uncertainty. The $\bar{{\mathcal M}}_{a=3}^{b=1}$ term $<0.001nb$ in this range, however, the $1MeV$ cross-section is slightly low. This suggests the $\bar{{\mathcal M}}_{a}^{b}$ for $b>1$ terms 
could be non-negligible at lower soft-photon cut-off values.}\footnote{For the higher mass regions in the  $e^{+}e^{-} \to \Upsilon(4s)  (m\gamma)\to BB(n\gamma)$ line scan, the truncation uncertainty 
for the expansion in 
 $\bar{{\mathcal M}}_{a}^{b=1}$ is estimated to be $\sim 0.5-1\%$.}.  
%Type IV \cite{Nugent:2022ayu} Exponentiation is
%applied for the initial-state and Model 2 is applied for the final-state YFS Exponentiation with a soft-photon cut-off of $1MeV$. 
The impact of the beam-spread is determined using numerical integration.   
Recently, an upgrade to SuperKEKB for polarization measurements at the BELLE-II experiment using a polarized $e^{-}$ beam has been proposed \cite{USBELLEIIGROUP:2022qro,Roney:2021pwz}. 
Therefore we present the angular distribution for unpolarized $e^{+}e^{-}$ beams and for polarized beams in Figure \ref{fig:spin}.
Figure \ref{fig:Fc} presents the relative enhancement of the  $e^{+}e^{-}\to  \Upsilon(4s) (m\gamma) \to B^{+}B^{-} (n\gamma)$ to the  $e^{+}e^{-}\to \Upsilon(4s) (m\gamma) \to B^{0}\bar{B}^{0}(n\gamma)$ cross-section due to the Coulomb potential in each of the final-state 
exponentiation models as a function of collision energy. From this it can be seen that
Model 1, yields an amplification of the $e^{+}e^{-}\to  \Upsilon(4s)  (m\gamma)\to B^{+}B^{-}(n\gamma)$ of $(20.28\pm0.10)\%$ relative to the $e^{+}e^{-}\to \Upsilon(4s) (m\gamma) \to B^{0}\bar{B}^{0}(n\gamma)$  cross-sections, where the theoretical uncertainties come from 
$\alpha(s)$ and the meson masses. This result is inconsistent 
with the world average branching fractions \cite{PDG2020}, and has a strong $\sqrt{s}$ dependence over the resonance which is inconsistent with the known line-shape as presented in \cite{PhysRevD72.032005,ARGUS:1994lbi}. 
However, when taking into account the impact of phase-space between the $e^{+}e^{-}\to \Upsilon(4s)  (m\gamma)\to B^{+}B^{-}(n\gamma)$ and $e^{+}e^{-}\to  \Upsilon(4s) (m\gamma) \to B^{0}\bar{B}^{0}(n\gamma)$ Models 2-4 are consistent with the 
world average branching fractions \cite{PDG2020}. An uncertainty of $\sim0.3-0.4\%$ is 
required to distinguish between Model 2, 3 and 4. This indicates that the Coulomb potential should be applied at either the quark or singlet level 
and not directly to the mesons (or baryons). 
This conclusion is also supported by the interpretation of the $e^{+}e^{-}\to p^{+}p^{-}$  results in \cite{BaldiniFerroli:2010ruh} using the \cite{BaBar:2005pon,BaBar:2007fsu} measurements. 
This has important implications for measurements of 
meson and baryon systems near threshold \footnote{The Coulomb potential is often taken into account at the meson level in $e^{+}e^{-}\to Hadron (\gamma)$ events used for determining the 
hadronic vacuum polarization \cite{MC:2010,PhysRevD.88.032013,Akhmetshin_2008,Lees_2013}.}. In particular this excludes the possibility of the Coulomb potential from explaining the low energy 
excess in the $\tau^{-}\to \pi^{-}\pi^{-}\pi^{+}\nu_{\tau}$ decay processes which was investigated in \cite{Nugent:2013hxa}.

\section{Conclusions}

The {\tt ee$\in$MC} generator has been extended to include $e^{+}e^{-} \to Hadrons (n\gamma)$ interactions, this includes the $e^{+}e^{-}\to \Upsilon(4s) (m\gamma) \to  B^{+}B^{-}(n\gamma)$, 
 $e^{+}e^{-}\to \Upsilon(4s) (m\gamma) \to  B^{0}\bar{B}^{0}(n\gamma)$ and $e^{+}e^{-}\to \pi^{-}\pi^{+}(n\gamma)$ interactions.
The $e^{+}e^{-}\to \pi^{-}\pi^{+}(n\gamma)$  is implemented within the  Gounaris-Sakurai \cite{Gounaris:1968mw} Model, while the  $e^{+}e^{-}\to \Upsilon(4s) (m\gamma) \to  B^{+}B^{-}(n\gamma)$
and  $e^{+}e^{-}\to \Upsilon(4s) (m\gamma) \to  B^{0}\bar{B}^{0}(n\gamma)$ are implemented within the Quark-Pair-Creation Model \cite{LeYaouanc:1972vsx,LeYaouanc:1977fsz,Ono:1980js,PhysRevD72.032005,ARGUS:1994lbi}.
Additional theoretical models and decay processes are expected to be implemented for $e^{+}e^{-} \to Hadrons (n\gamma)$ particularly those for studying the low mass range near or below 
the perturbative QCD threshold. These states form a complementary system for investigating the structure of low energy QCD to that of hadronic $\tau$ decays. Higher order radiative corrections 
are required within the YFS Exponentiation formalism 
to reach the theoretical precision required by initial-state-radiation measurements of the hadronic vacuum polarization at the B-Factories.
The impact of the Coulomb potential was investigated in terms of the radiative emission models in the $e^{+}e^{-} \to Hadrons (n\gamma)$ interactions, where the constrained phase-space in the
$e^{+}e^{-}\to \Upsilon(4s) (m\gamma) \to  B^{+}B^{-}(n\gamma)$, $e^{+}e^{-}\to \Upsilon(4s) (m\gamma) \to  B^{0}\bar{B}^{0}(n\gamma)$ processes was exploited. 
The impact of the final-state exponentiation models supports the assumption that the Coulomb potential should be applied at either the quark or singlet level in hadronic processes 
and not directly to the meson as is commonly done in the literature \cite{MC:2010,PhysRevD.88.032013,Akhmetshin_2008,Lees_2013}. This conclusion is consistent with \cite{BaldiniFerroli:2010ruh}.  
Moreover, this strongly suggests that the Coulomb potential 
is not a viable alternative explanation for the low mass excess in the $\tau^{-}\to \pi^{-}\pi^{-}\pi^{+}\nu_{\tau}$ decay processes \cite{Nugent:2013hxa} 
which tends to be interpreted in terms of a low mass hadronic scalar \cite{Nugent:2013hxa,CLEO3pi,Edwards:1999fj}.

\section*{Acknowledgement}
GCC Version 4.8.5 was used for compilation and the plots are generated using the external program GNUPlot  \cite{gnuplot4.2}.

\footnotesize
\bibliography{paper}
\normalsize

\noindent

\end{document}